\documentclass[a4paper,11pt]{article}
\pdfoutput=1 
\usepackage{jheppub} 
\usepackage[T1]{fontenc} 

\title{\boldmath Search for light mediators in the low-energy data of the CONNIE reactor neutrino experiment}

\author[a]{Alexis Aguilar-Arevalo,}
\author[b]{Xavier Bertou,}
\author[c]{Carla Bonifazi,}
\author[d]{Gustavo Cancelo,}
\author[a,1]{Brenda Aurea Cervantes-Vergara,\note{Corresponding author: bren.cv@ciencias.unam.mx}}
\author[e]{Claudio Chavez,}
\author[a]{Juan C. D'Olivo,}
\author[f]{Jo\~ao C. dos Anjos,}
\author[d]{Juan Estrada,}
\author[g]{Aldo R. Fernandes Neto,}
\author[d,h]{Guillermo Fernandez-Moroni,}
\author[c]{Ana Foguel,}
\author[d]{Richard Ford,}
\author[i]{Federico Izraelevitch,}
\author[j]{Ben Kilminster,}
\author[f]{H. P. Lima Jr,}
\author[f]{Martin Makler,}
\author[e]{Jorge Molina,}
\author[f]{Philipe Mota,}
\author[c]{Irina Nasteva,}
\author[h]{Eduardo Paolini,}
\author[e]{Carlos Romero,}
\author[a]{Youssef Sarkis,}
\author[b]{Miguel Sofo Haro,}
\author[d]{Javier Tiffenberg,}
\author[e]{and Christian Torres.}

\affiliation[a]{Universidad Nacional Aut\'onoma de M\'exico, Ciudad de M\'exico, M\'exico}
\affiliation[b]{Centro At\'omico Bariloche and Instituto Balseiro, Comisi\'on Nacional de Energ\'ia At\'omica (CNEA), Consejo Nacional de Investigaciones Cient\'ificas y T\'ecnicas (CONICET), Universidad Nacional de Cuyo (UNCUYO), San Carlos de Bariloche, Argentina.}
\affiliation[c]{Universidade Federal do Rio de Janeiro, Instituto de F\'isica, Rio de Janeiro, RJ, Brazil}
\affiliation[d]{Fermi National Accelerator Laboratory, Batavia, IL, United States}
\affiliation[e]{Facultad de Ingenier\'ia - Universidad Nacional de Asunci\'on, Asunci\'on, Paraguay}
\affiliation[f]{Centro Brasileiro de Pesquisas F\'isicas, Rio de Janeiro, RJ, Brazil}
\affiliation[g]{Centro Federal de Educa\c c\~ao Tecnol\'ogica Celso Suckow da Fonseca, Angra dos Reis, RJ, Brazil}
\affiliation[h]{Instituto de Investigaciones en Ingenier\'ia El\'ectrica, Departamento de Ingenier\'ia El\'ectrica y Computadoras, Universidad Nacional del Sur (UNS) - CONICET, Bah\'ia Blanca, Argentina}
\affiliation[i]{Universidad Nacional de San Mart\'in (UNSAM), Comisi\'on Nacional de Energ\'ia At\'omica (CNEA),  Consejo Nacional de Investigaciones Cient\'ificas y T\'ecnicas (CONICET), Argentina}
\affiliation[j]{Universit\"at Z\"urich Physik Institut, Zurich, Switzerland}

\collaboration{CONNIE Collaboration}

\abstract{The CONNIE experiment is located at a distance of 30 m from the core of a commercial nuclear reactor, and has collected a 3.7 kg-day exposure using a CCD detector array sensitive to an $\sim$1 keV threshold for the study of coherent neutrino-nucleus elastic scattering. Here we demonstrate the potential of this low-energy neutrino experiment as a probe for physics Beyond the Standard Model, by using the recently published results to constrain two simplified extensions of the Standard Model with light mediators. We compare the new limits with those obtained for the same models using neutrinos from the Spallation Neutron Source. Our new constraints represent the best limits for these simplified models among the experiments searching for CE$\nu$NS for a light vector mediator with mass $M_{Z^{\prime}}<$ 10 MeV, and for a light scalar mediator with mass $M_{\phi}<$ 30 MeV. These results constitute the first use of the CONNIE data as a probe for physics Beyond the Standard Model.}

\begin{document} 
\maketitle
\flushbottom

\section{Introduction} \label{sec:intro}

Coherent Elastic Neutrino-Nucleus Scattering (CE$\nu$NS) is a Standard Model (SM) process predicted more than 40 years ago~\cite{Freedman1974} through which a neutrino interacts coherently with all nucleons present in an atomic nucleus, resulting in an enhancement of the scattering cross section. The enhancement is approximately proportional to the square of the number of neutrons in the nucleus. However, despite its large cross section, this process took a long time to be observed due to the difficulty of measuring the low-energy nuclear recoils produced by the neutrino-nucleus scattering events. Recently, CE$\nu$NS was detected by the COHERENT collaboration~\cite{Coherent2017} thanks to the development of novel detectors and the unique neutrino beam facility of the Spallation Neutron Source (SNS) at the Oak Ridge National Laboratory.

The potential for CE$\nu$NS as a tool to search for beyond the Standard Model physics has been extensively discussed in the literature~\cite{Harnik2012,Billard2018}. More recently, data from COHERENT have opened a window into the low-energy neutrino sector, allowing them to impose new constraints on non-standard neutrino interactions (NSI)~\cite{Liao2017,Aristizabal2018,Khan2019} and to establish new limits on sterile neutrino models~\cite{Kosmas2017,Blanco2019}. Other searches for new physics with COHERENT data are discussed in \cite{Papoulias2018,Denton2018,Dutta2019,Miranda2019,Papoulias2019}.

The COHERENT experiment explores the high-energy tail of CE$\nu$NS, using spallation neutrinos with energies above 20 MeV in order to produce observable nuclear recoils in detectors with thresholds of the order of 20 keV. On the other hand, several efforts are ongoing to observe the CE$\nu$NS using neutrinos from nuclear reactors~\cite{Akimov2016,Agnolet2017,Ricochet2017,CONUS2019}, with typical neutrino energy of around 1 MeV. However, the SM signal has not yet been detected due to the very low-energy nuclear recoil signals produced.

The Coherent Neutrino-Nucleus Interaction Experiment (CONNIE)~\cite{CONNIEtalk2016} uses low-noise fully depleted charge-coupled devices (CCDs)~\cite{Holland2003} with the goal of measuring low-energy recoils from CE$\nu$NS produced by reactor antineutrinos with silicon nuclei~\cite{Guille2015}. The CONNIE engineering run, carried out in 2014--2015, is discussed in~\cite{CONNIE2016}. The detector installed in 2016 has an active mass of 73.2 g (12 CCDs) and is located 30 m from the core of the Angra 2 nuclear reactor, which has a thermal power of 3.95 GW. The CONNIE detector is sensitive to recoil energies down to 1 keV. A search for neutrino events is performed by comparing data with the reactor on (2.1 kg-day) and the reactor off (1.6 kg-day), the results show no excess in the reactor-on data~\cite{CONNIE2019}. A model independent 95\% Confidence Level (C.L.) upper limit for new physics was established at an event rate of $\sim$40 times the one expected from the SM at the lowest energies.

In this work we use the results recently published by CONNIE~\cite{CONNIE2019} to restrict the parameter space of two simplified extensions of the SM that have also been explored with the data of the COHERENT experiment. This approach shows the potential of experiments searching for CE$\nu$NS with low-energy reactor neutrinos to probe new physics in a way that is complementary to spallation neutrino experiments. The models we consider contain an additional light mediator: 1) a neutral vector boson $Z^{\prime}$, with mass $M_{Z^{\prime}}$, and 2) a scalar mediator $\phi$, with mass $M_{\phi}$. These simplified models represent a straightforward way to parametrize the reach for new physics in the low-energy neutrino sector, as discussed in Ref.~\cite{Cerdeno2016}. Testing such extensions of the SM is interesting since there are no constraints from the Large Hadron Collider (LHC) experiments when the mediator mass is below the GeV scale~\cite{Dutta2019}. From the theoretical side, these models have attracted considerable attention because, among other things, they connect to new ideas associated with sub-GeV dark matter in the range of MeV to GeV~\cite{Dutta2019,Essig2013}.

The paper is organized as follows. In Section~\ref{sec:cross}, we present the CE$\nu$NS cross section for the SM and its extensions with light mediators. In Section~\ref{sec:rate}, the event rate in CONNIE is calculated in these extended models. In Section~\ref{sec:limit}, the results of the CONNIE experiment are used to establish limits in the parameter space for the new vector and scalar bosons. Two appendices have been included with details of the neutrino flux and detector performance information needed to calculate the event rate. These details are published elsewhere and are given here for completeness.

\section{CE$\nu$NS cross section in the SM and its extensions}\label{sec:cross}

The CE$\nu$NS interaction happens when the three-momentum transfer $q = |\mathbf{q}|$ is small enough so that $q^2R^2<1$, with $R$ being the nuclear radius. In the laboratory frame, this corresponds to an energy of the incident antineutrino below 50 MeV. For reactor antineutrinos, the energies involved are below $\sim$5 MeV and the above condition is well satisfied. In the SM, the differential cross section for the coherent elastic scattering of antineutrinos off a nucleus at rest, with $Z$ protons, $N$ neutrons and mass $M$ is given by
\begin{equation}
\label{eq:crossSM}
\frac{d\sigma_{SM}}{dE_R}\left(E_{\bar{\nu}_e}\right)=\frac{G_F^2}{4\pi}Q_W^2\left(1-\frac{ME_R}{2E_{\bar{\nu}_e}^2}-\frac{E_R}{E_{\bar{\nu}_e}}+\frac{E_R^2}{2E_{\bar{\nu}_e}^2}\right)MF^2(q)\,,
\end{equation}
where $G_F$ is the Fermi coupling constant, $E_{\bar{\nu}_e}$ is the antineutrino energy, $E_R$ is the nuclear recoil energy and 
\begin{equation}
\label{eq:QW}
Q_W = N - (1 -4 \sin^2\theta_W)Z\,,
\end{equation}
is the weak nuclear charge. Here, $\theta_W$ is the weak mixing angle and $F(q)$ is the nuclear form factor, which can be expressed as~\cite{Klein2000}
\begin{equation}
\label{eq:ffact}
F(q) = \frac{4\pi \rho_0}{Aq^3}\left( \sin qR - qR \cos qR\right)
\frac{1}{1 + a^2q^2}\,,
\end{equation}
where $A$ is the atomic mass of the nucleus, $a=0.7 \times 10^{-13}$ cm is the range of the Yukawa potential considered, $R=r_0 A^{1/3}$ is the nuclear radius and $\rho_0~=~3/ 4\pi r_0^3$ is the nuclear density, with $r_0~=~1.3 \times 10^{-13}$ cm being the average radius of a proton in a nucleus. At low energies, $\sin^2\theta_W = 0.23857$~\cite{PDG}.

The SM cross-section is modified by the presence of new mediators, which couple to neutrinos and quarks. As mentioned before (Section \ref{sec:intro}), we consider here two simplified extensions of the SM with light mediators, which have been developed in Ref.~\cite{Cerdeno2016}. These extensions have recently 
been explored with data from the COHERENT experiment~\cite{Liao2017,Papoulias2018,Khan2019,Papoulias2019} and offer a good opportunity to demonstrate the complementarity of the two experimental approaches based on reactor and spallation neutrinos.

First, let us consider the non-standard interactions associated to a light vector mediator $Z^{\prime}$ with mass $M_{Z^{\prime}}$ and coupling $g^{\prime}$. To keep things as simple as possible, we assume that there is no $Z$-$Z'$ mixing and that the $Z^{\prime}$ has a purely vector interaction with the fermions of the SM, with a universal flavor-conserving coupling to the first generation of quarks and leptons. At tree level, the net effect on the CE$\nu$NS is merely a modification of the global factor $Q_W^2$ in Eq.~\eqref{eq:crossLV} ~\cite{Liao2017}. Thus, the differential cross section now becomes:
\begin{equation} 
\label{eq:crossLV}
\frac{d\sigma_{SM+Z^{\prime}}}{dE_R}\left(E_{\bar{\nu}_e}\right)=\left(1-\frac{Q_{Z^{\prime}}}{Q_W}\right)^2\frac{d\sigma_{SM}}{dE_R}\left(E_{\bar{\nu}_e}\right)\,,
\end{equation}
where $d\sigma_{SM}/dE_R$ is given in Eq.~\eqref{eq:crossSM} and
\begin{equation}
\label{eq:QLV}
Q_{Z^{\prime}} = \frac{3\left(N+Z\right){g^{\prime}}^2}{\sqrt{2}G_F\left(2M E_R + M_{Z^{\prime}}^2\right)}\,.
\end{equation}

To probe the new interactions related to a light scalar mediator $\phi$, with mass $M_{\phi}$, we adopt a simplified model in which the couplings to quarks are all the same~\cite{Cerdeno2016}. Since Yukawa-like interactions change the chirality of the particles involved, there is no contribution to the transition probability coming from the interference with the chirality-preserving $Z$-boson interactions~\cite{Farzan2018}. Then, the differential cross-section can be written as the sum of the SM contribution plus the one of the new light scalar:
\begin{equation}
\label{eq:crossLS}
\frac{d\sigma_{SM+\phi}}{dE_R}(E_{\bar{\nu}_e})~=~\frac{d\sigma_{SM}}{dE_R}(E_{\bar{\nu}_e})+ \frac{d\sigma_{\phi}}{dE_R}(E_{\bar{\nu}_e})\,,
\end{equation}
where
\begin{equation}
\frac{d\sigma_{\phi}}{dE_R}(E_{\bar{\nu}_e})~=~\frac{G_F^2}{4\pi}Q_{\phi}^2\left(\frac{2ME_R}{E_{\bar{\nu}_e}^2}\right)MF^2(q)\,,
\end{equation}
with
\begin{equation}
\label{eq:QLS}
Q_{\phi}~=~\frac{\left(14 N + 15.1 Z\right)g_{\phi}^2}{\sqrt{2}G_F\big(2ME_R + M_{\phi}^2\big)}\,.
\end{equation}
Here, $g_{\phi}^2 \equiv g_{\nu} g_q$ where $g_{\nu}$ is the neutrino coupling and $g_q$ is the common coupling to quarks.
Light dark matter models that thermalize through the Higgs portal require a light scalar mediator which mixes with the Higgs boson~\cite{Krnjaic2016}. As a consequence, the scalar mediator acquires a coupling to the SM fermions $g_f = m_f \sin \theta/v$, where $m_f$ is the fermion mass, $v$ is the vacuum expectation value of the Higgs field ($v \sim 246$ GeV) and $\sin{\theta}$ parametrizes the mixing between the dark sector and the SM. In this case,
\begin{equation}
\label{eq:Hportal}
    g_{\phi}^2\simeq \frac{ m_{\nu} m_{_{\!\mathcal N}}}{v^2}\sin^2{\theta}\left(\frac{0.168 N + 0.164 Z}{14 N + 15.1 Z}\right),
\end{equation}
where $m_{\nu}$ is the mass of the neutrino and $m_{_{\!\mathcal N}}\!=938. 9$ MeV is the average mass of nucleons. The numerical factors 0.168 and 0.164 were computed using the values of the coefficients $f^{(n,p)}_{Tq}$ in Ref.~\cite{Ellis2000} that are consistent with Ref.~\cite{Cerdeno2016}. In this way, constraints on $g_{\phi}^2$ imposed in the simplified model can be mapped to constraints on $m_{\nu}\sin^2{\theta}$ in the Higgs portal model using Eq.~\eqref{eq:Hportal}.

In the next section, the formulae for the cross-sections given above are used to compute the event rate in the CONNIE experiment as a function of the parameters (coupling and mass) of each model. We do so for silicon nuclei ($N~=~Z~=~14$), in which case Eq.~\eqref{eq:QW} reduces to $Q_W~=~56\sin^2\theta_W$, while in Eqs.~\eqref{eq:QLV} and~\eqref{eq:QLS} we have $3(N+Z)~=~84$ and $14N+15.1Z~=~407.4$, respectively.

\section{Event rate in CONNIE} \label{sec:rate}

The differential event rate as a function of the nuclear recoil energy $E_R$ in the CONNIE experiment is
\begin{equation}
\frac{dR}{dE_R}= N_T\int_{E^{\rm min}_{\bar{\nu}_e}}^{\infty} \frac{d\Phi}{dE_{\bar{\nu}_e}}\,\frac{d\sigma}{dE_R}\, dE_{\bar{\nu}_e}\,,
\end{equation}
where $d\sigma/dE_R$ is the CE$\nu$NS differential cross section (Eq.~\eqref{eq:crossSM},~\eqref{eq:crossLV} or~\eqref{eq:crossLS}), $d\Phi/dE_{\bar{\nu}_e}$ is the reactor antineutrino flux as a function of the energy (discussed in Appendix~\ref{app:flux}), $N_T$ is the number of nuclei in the detector and $E^{\rm min}_{\bar{\nu}_e}~=~\Big(E_R + \sqrt{E_R^2 + 2ME_R}\Big)/2$ is the minimal antineutrino energy that can produce a nuclear recoil with energy $E_R$.

The CONNIE sensors detect the ionization produced by the recoiling silicon nuclei. As discussed in Appendix~\ref{app:fits}, the quenching factor $Q$ relates the ionizing energy $E_I$ to the recoil energy: $E_R~=~E_I/Q(E_I)$. Taking this relation into account we get the differential event rate as a function of $E_I$,

\begin{equation} \label{eq:difrateEI}
\frac{dR}{dE_I}=\frac{dR}{dE_R}\frac{dE_R}{dE_I} =\frac{dR}{dE_R} \frac{1}{Q} \left(1-\frac{E_I}{Q}\frac{dQ}{dE_I} \right)\,.
\end{equation}

The total rate $R$ in the CONNIE experiment is given by
\begin{equation}
R =\int_{E_{\rm th}}^{\infty} \epsilon(E_M)\,\frac{dR}{dE_M}\, dE_M\,,
\end{equation} 
where $E_M$ is the measured energy, $\epsilon(E_M)$ denotes the reconstruction efficiency (see Appendix~\ref{app:fits}) and $E_{\rm th} = 0.064$ keV is the detector threshold given by the efficiency curve. Assuming a Gaussian detector response, the differential event rate as a function of $E_M$ is
\begin{equation} \label{eq:difrateEM}
\frac{dR}{dE_M} = \frac{{\int_0^{\infty} G\left(E_M, E_I; \sigma^2_I\right)}\,\dfrac{dR}{dE_I}\, dE_I}{\int_0^{\infty}G\left(E_M, E_I; \sigma^2_I\right) dE_I}\,,
\end{equation}
where
\begin{equation} \label{eq:res}
G\left(E_M, E_I; \sigma^2_I\right) = \frac{1}{\sqrt{2 \pi \sigma^2_I}}\exp \left\{-\dfrac{\left(E_M-E_I\right)^2}{2\sigma^2_I}\right\}\,,
\end{equation}
with $\sigma^2_I=\left(0.034 \textrm{ keV}\right)^2 + F E_{\rm eh} E_I$ characterizing the energy resolution of a typical CONNIE CCD~\cite{DAMIC2016}. Here, $F$ is the Fano factor (0.133) and $E_{\rm eh}$ is the mean ionization energy required for photons to produce an electron-hole pair in silicon (0.003745 keV)~\cite{Ryan1973}.

\section{Search for light mediators using CONNIE data} \label{sec:limit}

\begin{figure}[ht]
\centering
 \includegraphics[scale=0.5]{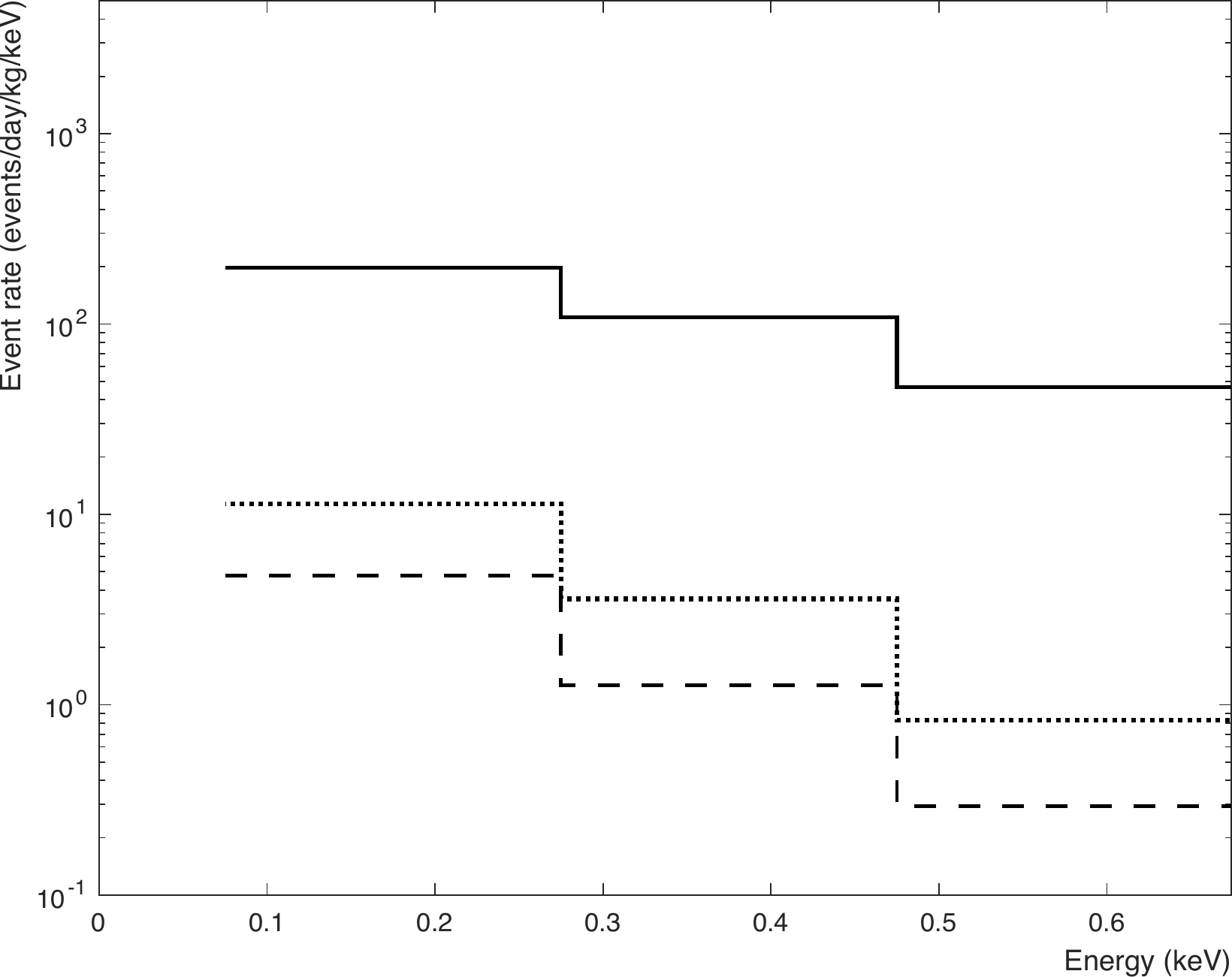}
\caption{95\% confidence level limits from RON-ROFF measurements (solid line) and CE$\nu$NS expected event rate using measurements in Ref.~\cite{Chavarria2016} (dashed line) and expressions in Ref.~\cite{Lindhard1963} (dotted line) for the quenching factor. Figure from Ref.~\cite{CONNIE2019}.}
\label{fig:resultsconnie}
\end{figure}

The results in the recent CONNIE run discussed in Ref.~\cite{CONNIE2019} establish new limits for the rate of low-energy antineutrino events coming from non-standard interactions $R_{NSI}$ based on a comparison of reactor-on (RON) and reactor-off (ROFF) data. The limits are calculated assuming a Poisson distribution for the binned energy spectrum obtained in RON and ROFF conditions. The excess in RON events is calculated as the difference RON-ROFF, consistent with zero. Assuming this difference as gaussian distributed with standard deviation determined from the uncertainties in RON and ROFF, the 95\% C.L.~upper limits are established using the frequentist method for determining a one sided confidence interval for gaussian variable. (See Ref.~\cite{CONNIE2019} for a complete description of the analysis.) These limits are shown in Fig.~\ref{fig:resultsconnie}, together with the expected event rate for the Standard Model, $R_{SM}$.

From the lowest-energy bin of this figure, we see that the 95\% C.L. upper limit established by CONNIE is 41 times above the SM prediction using the quenching factor measurements in Ref.~\cite{Chavarria2016}. This allows us to set the limit $R_{NSI} < R_{SM}\times 41$ for 0.075 keV $ < E_M < $ 0.275 keV (or recoil energies 0.784 keV $< E_R <$ 1.834 keV). This limit is presented in Ref.~\cite{CONNIE2019} as a model independent limit for a counting experiment in the lowest energy bin for the data in CONNIE. Here, we study this limit in the context of light mediators, as discussed in Section~\ref{sec:cross}. The upper limits resulting from higher energies are weaker because the SM prediction drops fast as the energy increases. For the next energy bin, 0.275 keV $< E_M <$ 0.475 keV, the limit is $R_{NSI} < R_{SM}\times 84$. For this reason we perform the search for the two extensions of the SM described in Section \ref{sec:cross}, using only the lowest-energy bin of the limits published by the CONNIE Collaboration in Ref.~\cite{CONNIE2019}. We leave for future work an analysis using the full spectral shape of the reactor data from CONNIE.

The 95\% C.L.~exclusion limits for the parameters of the simplified models with a light mediator are calculated as the curve in the 2D parameter space for which the rate of non-standard interactions is
\begin{equation}
R_{NSI}( M , g ) = \int_{E_1}^{E_2} \epsilon(E_M)\,\frac{dR_{NSI}}{dE_M}\, dE_M\, = 41~R_{SM} \,,
\end{equation} 
where $M$ and $g$ are $M_{Z^{\prime}} (M_{\phi})$ and $g^{\prime} (g_\phi)$ for the vector(scalar) mediator models. $E_1 = 0.075$ keV and $E_2 = 0.275$ keV are determined by the lowest energy bin in the CONNIE data, $dR_{NSI}/dE_M$ is calculated using Eq.~\eqref{eq:difrateEM}, considering the non-standard differential cross sections from Eqs.~\eqref{eq:crossLV} and ~\eqref{eq:crossLS}, and $R_{SM}$ is calculated in a similar way, using the SM differential cross section in Eq.~\eqref{eq:crossSM}.

\begin{figure}[ht] 
\centering
\includegraphics[scale=1]{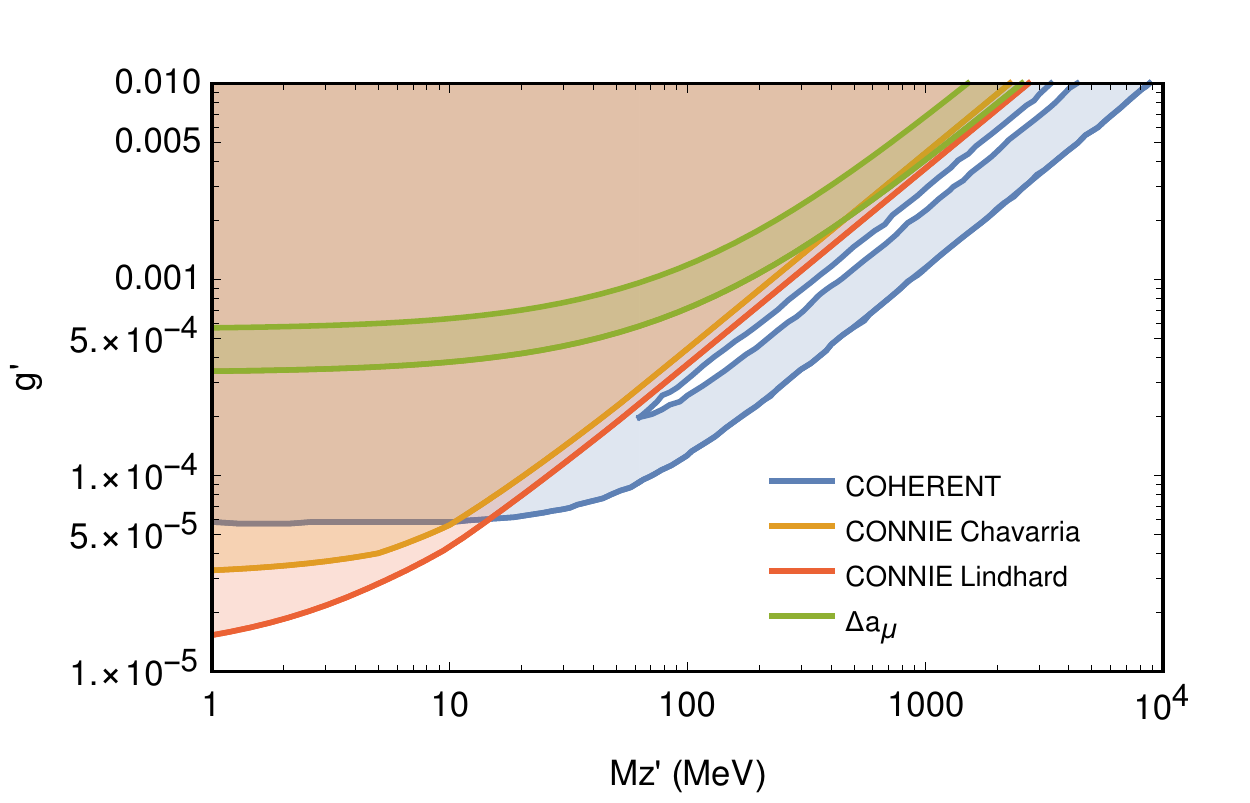}
\caption{Exclusion region in the $\left(M_{Z^{\prime}}, g'\right)$ plane from the CONNIE results assuming quenching given by the fit to the measurements in Ref.~\cite{Chavarria2016} (orange) and the expressions in Ref.~\cite{Lindhard1963} (red). The COHERENT limit curve~\cite{Liao2017} (blue) and the 2$\sigma$ allowed region to explain the anomalous magnetic moment of the muon $(\Delta a_{\mu}=268 \pm 63 \times 10^{-11})$~\cite{Jegerlehner2009,Krasnikov2017} (green) are shown for reference.} \label{fig:exclusion}
\end{figure}

\begin{figure}[ht] 
\centering
\includegraphics[scale=1]{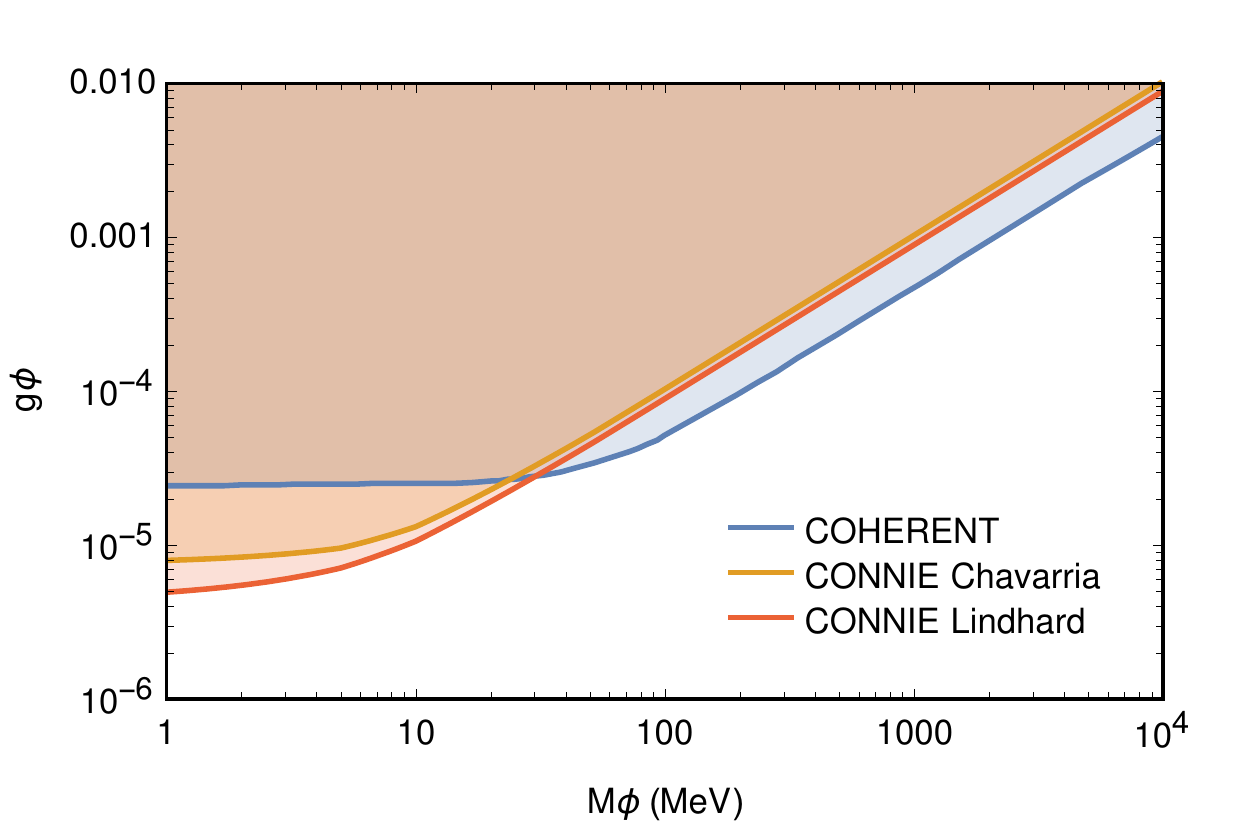}
\caption{Exclusion region in the $\left(M_{\phi}, g_{\phi}\right)$ plane from the CONNIE results assuming quenching given by the fit to the measurements in Ref.~\cite{Chavarria2016} (orange) and the expressions in Ref.~\cite{Lindhard1963} (red). The 90\% COHERENT limit curve in Ref.~\cite{Khan2019} (blue) is shown for reference.} \label{fig:exclusionLS}
\end{figure}

The resulting limits are shown in Figs.~\ref{fig:exclusion} and~\ref{fig:exclusionLS}. The most significant systematic uncertainty in these limits comes from the quenching factor measurement, as discussed in Ref.~\cite{CONNIE2019}. The most used model for the quenching factor in literature is the Lindhard model~\cite{Lindhard1963}. There is only one measurement performed for nuclear recoils in silicon at low energies using CCDs similar to those in CONNIE~\cite{Chavarria2016}. The results from this measurement are not consistent with the Lindhard model. To quantify the systematic uncertainty associated with the quenching factor, we also include in Figs.~\ref{fig:exclusion} and~\ref{fig:exclusionLS} the exclusion region calculated using the quenching factor from Ref.~\cite{Lindhard1963}. Additional systematic uncertainties for the limit established in Ref.~\cite{CONNIE2019} are sub-dominant compared to the effect of the quenching factor. These uncertainties are related to the reactor flux, the detection efficiency and the stability in the detector energy calibration, contributing less than 5\%, 10\% and 2\%, respectively, to the event rate in the lowest-energy bin. These lower level effects will become relevant once the uncertainty in the quenching factor is significantly reduced in future analyses.

Regarding the simplified vector mediator model, according to the expression in Eq.~\eqref{eq:crossLV}, the contribution of the additional mediator to the event rate is proportional to ${g^{\prime}}^2/(2M E_R + M_{Z^{\prime}}^2)$. Therefore, for a light mediator, $M_{Z^{\prime}} \ll \sqrt{2M E_R}$, the NSI contribution depends only on $g^{\prime}$ and the limit becomes independent of mass. For a heavy mediator, $M_{Z^{\prime}} \gg \sqrt{2ME_R}$, the NSI rate contribution is proportional to the ratio $g^{\prime}/M_{Z^{\prime}}$. These two regimes are visible in Fig.~\ref{fig:exclusion}. Moreover, the CONNIE limit curve confirms the statement in Ref.~\cite{Liao2017}, disfavoring a light vector mediator to explain the discrepancy in the anomalous magnetic moment of the muon.

The simplified scalar mediator model exhibits a similar behavior. In this case, according to the expression in Eq.~\eqref{eq:crossLS}, the contribution of the additional mediator to the event rate is proportional to $g_{\phi}^2/(2M E_R + M_{\phi}^2)$. Again, for a light mediator the NSI contribution depends only on the coupling and for a heavy mediator it depends on the ratio $g_{\phi}/M_{\phi}$. These two cases are readily seen in Fig.~\ref{fig:exclusionLS}.

\section{Conclusion}

We use the recent results of the CONNIE experiment~\cite{CONNIE2019} to determine the constraints on neutrino neutral-current interactions mediated by a light vector-boson and a light scalar simplified extensions of the SM. These models were recently developed to search for new physics in the low-energy neutrino sector~\cite{Cerdeno2016}. Our analysis produces the best limits among the experiments searching for CE$\nu$NS in the low-mass regime, $M_{Z^{\prime}}< 10$~MeV in the case of a vector and $M_{\phi}< 30$~MeV in the case of a scalar, extending beyond the region excluded by the COHERENT results shown in Refs.~\cite{Liao2017, Khan2019}. With some additional assumptions, the limits coming from experiments searching for CE$\nu$NS can be compared to other limits coming from different experiments, as is shown in Refs.~\cite{Dent2017,Farzan2018,Abdullah2018}. The results presented here constitute the first search for NSI with reactor neutrinos and CCDs, and are expected to be the first in a series of searches using the CONNIE data.

For large mediator masses, the limits established by CONNIE are less stringent than those from COHERENT. This is related to the fact that CONNIE is looking at a flux of lower-energy antineutrinos. Higher-energy nuclear recoils produced by neutrinos from the SNS give access to the structure function in Eq.~\eqref{eq:ffact}, which allows to explore nuclear physics, and have the unique advantage of timing discussed in Ref.~\cite{Dutta2019}. However, the results coming from neutrinos from nuclear reactors have the advantage of being practically free of the uncertainties coming from the not well known nuclear structure, as the form factor in Eq.~\eqref{eq:ffact} can be approximated to unity. Moreover, because of their lower energy thresholds, reactor experiments provide a more powerful probe into new physics at low energies, such as the signatures expected for the simplified light mediators models. This makes evident the complementarity of two different techniques to explore new physics with neutrinos from reactor (CONNIE) and from the Spallation Neutron Source (COHERENT).

Two features of the present study are expected to be improved in the future. The current analysis is based on a counting experiment, comparing the number of events above threshold in CONNIE with the expectations from two simplified models with light mediators. More powerful limits are expected when spectral information of the CONNIE data is included in the analysis. Additionally, the CONNIE collaboration has recently modified the operation of the detector, performing hardware binning on the CCDs (adding charge of several pixels before readout) and reducing the effect of readout noise for the low-energy events. This operating mode improves the efficiency of the detector at low energies. Updated results using data taken with this mode are expected soon.

Last but not least, the quenching factor for nuclear recoils is the dominant systematic uncertainty for the limit to NSI in Ref.~\cite{CONNIE2019}, which is the basis of the analysis presented here. Significant effort is ongoing to improve our understanding of this important quantity at low energies, and it will be critical for any future experiment using silicon at even lower energies than CONNIE.

\appendix


\section{Reactor flux} \label{app:flux}
The CONNIE detector is located 30 m from the core of the Angra 2 reactor of the Almirante Alvaro Alberto Nuclear Power Plant, in Rio de Janeiro, Brazil. The thermal power of this reactor is 3.95~GW = $2.46 \times 10^{22}$ MeV/s. Considering that the average energy released per fission is 205.24 MeV, the number of fissions per second for this reactor is $n_f = 1.2 \times 10^{20}$.

The $\beta$ decays of the fission products, following the fission of four principal fissile isotopes ($^{235}$U, $^{238}$U, $^{239}$Pu and $^{241}$Pu), produce a large number of $\bar{\nu}_e$, contributing approximately 84\% to the total reactor's antineutrino flux. Each fissile isotope has its own $\bar{\nu}_e$ spectrum, which has been taken from~\cite{Vogel1989}. For energies below 2 MeV, the antineutrino spectra are given as tabulated values in Table~\ref{tab:reactortabval},
\begin{table}[ht]
\begin{center}
\begin{tabular}{|c|c|c|c|c|}
\hline
E$_{\bar{\nu_{e}}}$(MeV) & $^{235}$U & $^{239}$Pu & $^{238}$U & $^{241}$Pu\\
\hline
7.813 $\times 10^{-3}$ & 0.024 & 0.14 & 0.089 & 0.20\\
1.563 $\times 10^{-2}$ & 0.092 & 0.56 & 0.35 & 0.79\\
3.12 $\times 10^{-2}$ & 0.35 & 2.13 & 1.32 & 3.00\\
6.25 $\times 10^{-2}$ & 0.61 & 0.64 & 0.65 & 0.59\\
0.125 & 1.98 & 1.99 & 2.02 & 1.85\\
0.25 & 2.16 & 2.08 & 2.18 & 2.14\\
0.50 & 2.66 & 2.63 & 2.91 & 2.82\\
0.75 & 2.66 & 2.58 & 2.96 & 2.90\\
1.0 & 2.41 & 2.32 & 2.75 & 2.63\\
1.5 & 1.69 & 1.48 & 1.97 & 1.75\\
2.0 & 1.26 & 1.08 & 1.50 & 1.32\\
\hline
\end{tabular}
\end{center}
\caption{Tabulated values of the antineutrino spectrum of each fissile isotope in units of $\bar{\nu_{e}}$/MeV/fis.} \label{tab:reactortabval}
\end{table}
while for energies above 2 MeV these spectra are described by the parametric expression
\begin{equation}
\frac{dN_{\bar{\nu}_e}}{dE_{\bar{\nu}_e}} = a e^{a_0 + a_1 E_{\bar{\nu}_e} + a_2 E_{\bar{\nu}_e}^2}\,,
\end{equation}
where the fitted parameters are listed in Table~\ref{tab:reactorfitval}.
\begin{table}[ht]
\begin{center}
\begin{tabular}{|c|c|c|c|c|}
\hline
Parameter & $^{235}$U & $^{239}$Pu & $^{238}$U & $^{241}$Pu\\
\hline
a & 1.0461 & 1.0527 & 1.0719 & 1.0818\\
a$_0$ & 0.870 & 0.896 & 0.976 & 0.793\\
a$_1$ & $-$0.160 & $-$0.239 & $-$0.162 & $-$0.080\\
a$_2$ & $-$0.0910 & $-$0.0981 & $-$0.0790 & $-$0.1085\\
\hline
\end{tabular}
\end{center}
\caption{Fitted parameters of the antineutrino spectrum of each fissile isotope.} \label{tab:reactorfitval}
\end{table}

Another process that contributes approximately 16\% to the reactor antineutrino flux is the neutron capture of $^{238}$U nuclei. These nuclei capture 0.6 neutrons per fission via the reaction $^{238}$U + n $\rightarrow ^{239}$U $\rightarrow ^{239}$Np $\rightarrow ^{239}$ Pu. The $\beta$ decay of $^{239}$U produces two $\bar{\nu}_e$. The antineutrino spectrum of this process was extracted from~\cite{Texono2007}.

Each process previously considered has its own $\bar{\nu}_e$ yield and fission rate. The respective values were taken from~\cite{Texono2007} and are shown in Table~\ref{tab:rates}.
\begin{table}[ht]
\begin{center}
\begin{tabular}{|c|c|c|c|}
\hline
Process & Relative rate per fission & $N_{\bar{\nu}_e}$ per process & $N_{\bar{\nu}_e}$ per fission\\
\hline
$^{235}$U fission & 0.55 & 6.14 & 3.4\\
$^{239}$Pu fission & 0.32 & 5.58 & 1.8\\
$^{238}$U fission & 0.07 & 7.08 & 0.5\\
$^{241}$Pu fission & 0.06 & 6.42 & 0.4\\
$^{238}$U$(n,\gamma)^{239}$U & 0.60 & 2.00 & 1.2\\
\hline
\end{tabular}
\end{center}
\caption{Typical relative fission contribution and $\bar{\nu}_e$ yield for each process considered \cite{Texono2007}.} \label{tab:rates}
\end{table}

In order to obtain the total antineutrino reactor energy spectrum per fission, $d\mathcal{N}_{\bar{\nu}_e}/dE_{\bar{\nu}_e}$, the individual spectra were summed after being normalized and multiplied by their corresponding $\bar{\nu}_e$ yield per fission. 

\begin{figure}[ht]
\centering
\includegraphics[scale=0.8]{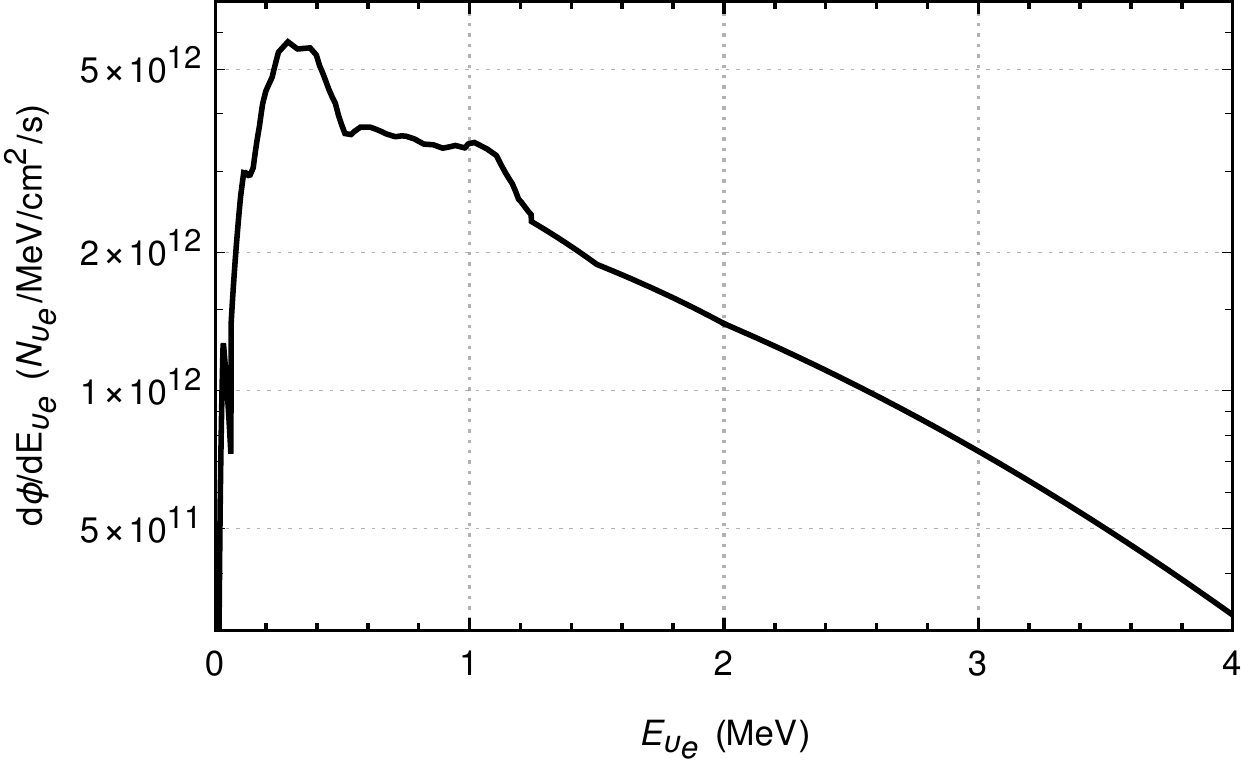}
\caption{Antineutrino flux expected at the CONNIE detector.} \label{fig:flux}
\end{figure}

The total antineutrino flux as function of the energy at the CONNIE detector, in units of number of antineutrinos per MeV per cm$^2$ per second, is given by
\begin{equation}
\frac{d\Phi}{dE_{\bar{\nu}_e}} = \frac{n_f}{4\pi d^2} \left(\frac{d\mathcal{N}_{\bar{\nu}_e}}{dE_{\bar{\nu}_e}}\right)\,,
\end{equation}
where $d=30\times10^{2}$ cm is the distance between the reactor and the detector. The flux expected at the CONNIE detector is shown in Fig.~\ref{fig:flux}.

\section{Fitting functions for quenching factor and efficiency} \label{app:fits}

When a nuclear recoil is produced inside the detector, a part of its energy generates charge carriers ($E_I$) and the rest contributes to the increase of the thermal energy of the system. The nuclear recoil quenching factor $Q$ is defined as the fraction of the total recoil energy $E_R$ that is used to produce ionization
\begin{equation}
Q=E_I/E_R\,.
\end{equation}

\begin{figure}[ht]
\centering
\includegraphics[scale=0.8]{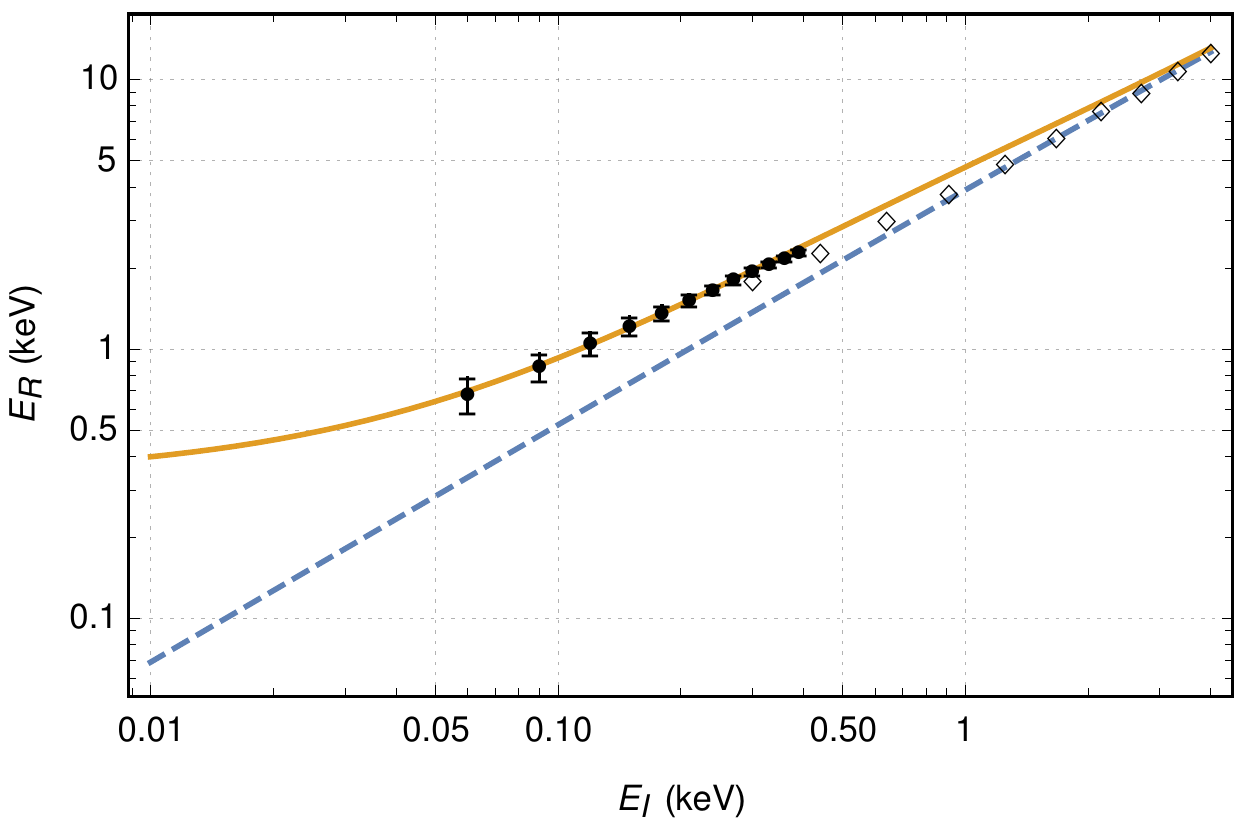}
\caption{Nuclear recoil quenching factors: Lindhard~\cite{Lindhard1963} (dashed blue line), measurements in~\cite{Izraelevitch2017} ($\diamond$) and in~\cite{Chavarria2016} ($\bullet$) and fit described by Eq.~\eqref{eq:fitChav} (solid orange line).}\label{fig:quench}
\end{figure}

For $E_R \gtrsim 4$ keV, the nuclear recoil quenching factor is well modelled by Lindhard~\cite{Lindhard1963}. Two measurements of the quenching factor for $E_R$ below 4 keV were performed using similar detectors in different experiments~\cite{Chavarria2016,Izraelevitch2017}. An analytical fit to the measurements in~\cite{Chavarria2016} is used here, parametrized as
\begin{equation} \label{eq:fitChav}
Q(E_I)=\frac{p_3 E_I +  p_4 E_I^2 + E_I^3}{p_0 + p_1 E_I + p_2 E_I^2}\,,
\end{equation}
with $p_0 = 56$ keV$^3$, $p_1 = 1096$ keV$^2$, $p_2 = 382$ keV, $p_3 = 168$ keV$^2$ and $p_4 = 155$ keV, as shown in Fig.~\ref{fig:quench}.

\begin{figure}[ht]
\centering
\includegraphics[scale=0.8]{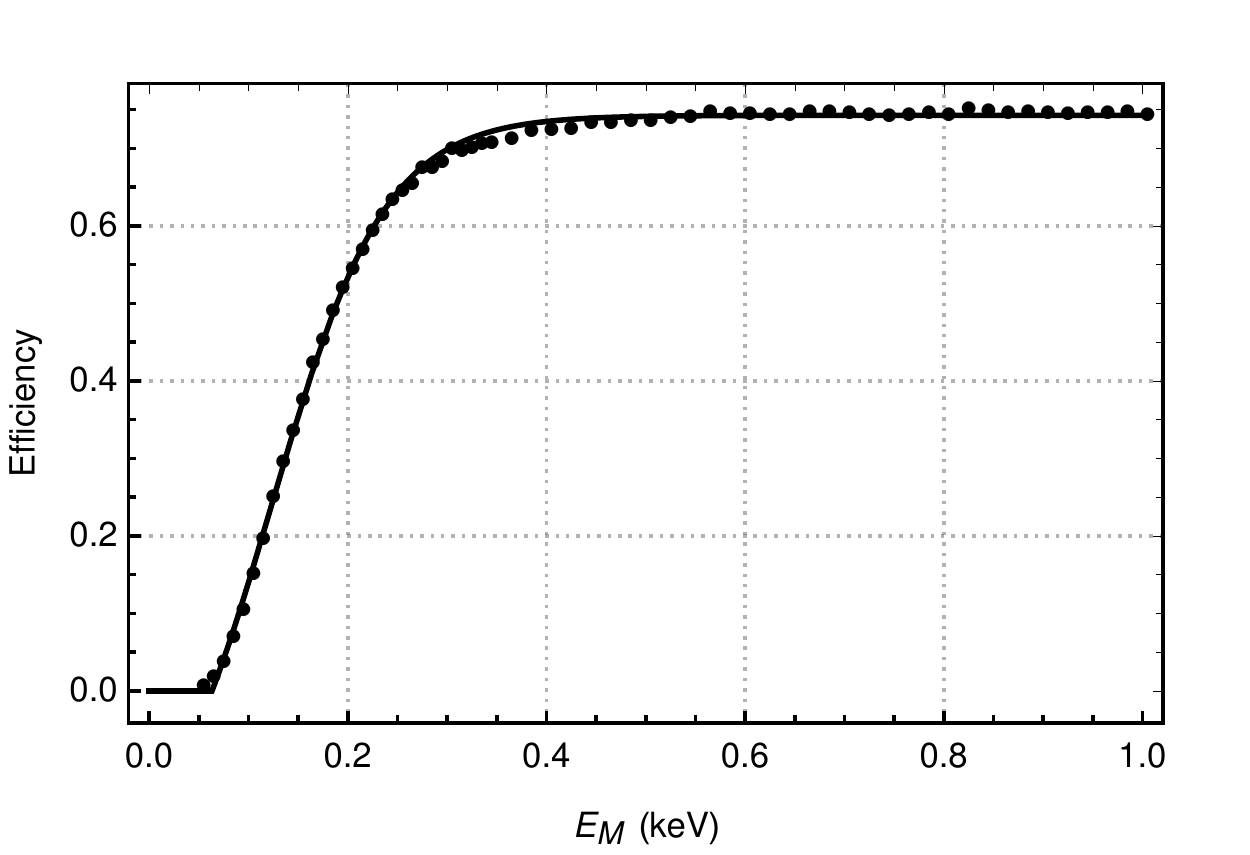}
\caption{Processing efficiency for CONNIE data obtained from simulated events~\cite{CONNIE2019} ($\bullet$) and the fit described by Eq.~\eqref{eq:eff} (solid line).}\label{fig:eff}
\end{figure}

In order to extract and reconstruct the events registered during a CCD exposure, a set of processing tools is used. The reconstruction efficiency for these tools has been evaluated in Ref.~\cite{CONNIE2019} using simulated events. The computed efficiency is fitted well for $E_M > 64$ eV by
\begin{equation} \label{eq:eff}
\epsilon(E_M) = b - \left[1+e^{b_0(E_M-b_1)}\right]^{-1}\,,
\end{equation}
where $b = 0.7426$, $b_0 = 17.47$ and $b_1 = 0.1239$, as shown in Fig.~\ref{fig:eff}.

\acknowledgments
We thank Eletrobras Eletronuclear for access to the Angra 2 reactor site and for the support of their personnel, in particular Ilson Soares and Gustavo Coelho, to the CONNIE activities. We thank the Silicon Detector Facility team at Fermi National Accelerator Laboratory for being the host lab for the assembly and testing of the detectors components used in the CONNIE experiment. We acknowledge the support from the former Brazilian Ministry for Science, Technology, and Innovation (currently MCTIC), the Brazilian Center for Physics Research and the Brazilian funding agencies FAPERJ (grants E-26/110.145/2013, E-26/210.151/2016), CNPq, and FINEP (RENAFAE grant 01.10.0462.00); and M\'exico's CONACYT (grant No.~240666) and DGAPA-UNAM (PAPIIT grant IN108917).

\bibliography{References}

\end{document}